\documentclass[12pt,preprint]{aastex}
\usepackage{xcolor}
\usepackage{soul}
\usepackage{color}
\usepackage{epsfig}
\usepackage{stix}

\shorttitle{The 'Cosmic Seagull': a highly magnified disk-like galaxy at $z\simeq2.8$ behind the Bullet Cluster}
\shortauthors{Motta et al.}

\begin{document}

\title{The `Cosmic Seagull': a highly magnified disk-like galaxy at $z\simeq2.8$ behind the Bullet Cluster}

\author{V. Motta} 
\affil{Instituto de F\'{\i}sica y Astronom\'{\i}a, Universidad de Valpara\'{\i}so, Avda. Gran Breta\~na 1111, Playa Ancha, Valpara\'{\i}so 2360102, Chile.}

\author{E. Ibar}
\affil{Instituto de F\'{\i}sica y Astronom\'{\i}a, Universidad de Valpara\'{\i}so, Avda. Gran Breta\~na 1111, Playa Ancha, Valpara\'{\i}so 2360102, Chile.}

\author{T. Verdugo}
\affil{Instituto de Astronom\'ia, Universidad Nacional Aut\'onoma de M\'exico, Apartado postal 106, C.P. 22800,  Ensenada, B.C., M\'exico.}

\author{J. Molina}
\affil{Departamento de Astronom\'{\i}a, Universidad de Chile, Casilla 36-D, Santiago, Chile.}

\author{T.M. Hughes}
\affil{Instituto de F\'{\i}sica y Astronom\'{\i}a, Universidad de Valpara\'{\i}so, Avda. Gran Breta\~na 1111, Playa Ancha, Valpara\'{\i}so 2360102, Chile.}
\affil{CAS Key Laboratory for Research in Galaxies and Cosmology,
Department of Astronomy, University of Science and Technology of
China, Hefei 230026, China.}
\affil{School of Astronomy and Space Science, University of Science and
Technology of China, Hefei 230026, China.}
\affil{Chinese Academy of Sciences South America Center for Astronomy,
China-Chile Joint Center for Astronomy, Camino El Observatorio 1515,
Las Condes, Santiago, Chile.}

\author{M. Birkinshaw}
\affil{University of Bristol, HH Wills Physics Laboratory, Tyndall Avenue, Bristol BS8 1TL, United Kingdom.}

\author{O. L\'opez-Cruz}
\affil{Instituto Nacional de Astrof\'{\i}sica, \'Optica y Electr\'onica (INAOE), Coordinaci\'on de Astrof\'{\i}sica, Luis Enrique Erro No. 1, Tonantzintla, Puebla, C.P. 72840, M\'exico.}

\author{J.H. Black}
\affil{Department of Space, Earth and Environment, Chalmers University of Technology, Onsala Space Observatory, 43992, Onsala, Sweden.}

\author{D. Gunawan}
\affil{Instituto de F\'{\i}sica y Astronom\'{\i}a, Universidad de Valpara\'{\i}so, Avda. Gran Breta\~na 1111, Playa Ancha, Valpara\'{\i}so 2360102, Chile.}

\author{C. Horellou}
\affil{Department of Space, Earth and Environment, Chalmers University of Technology, Onsala Space Observatory, 43992, Onsala, Sweden.}

\author{J. Maga\~na}
\affil{Instituto de F\'{\i}sica y Astronom\'{\i}a, Universidad de Valpara\'{\i}so, Avda. Gran Breta\~na 1111, Playa Ancha, Valpara\'{\i}so 2360102, Chile.}

\begin{abstract}
We present Atacama Large Millimeter/submillimeter Array measurements of the `Cosmic Seagull', a strongly magnified galaxy at $z=2.7779$ behind the Bullet Cluster. We report CO(3-2) and continuum 344~$\mu$m (rest-frame) data at one of the highest differential magnifications ever recorded at submillimeter wavelengths ($\mu$ up to $\sim50$), facilitating a characterization of the kinematics of a rotational curve in great detail (at $\sim 620$~pc resolution in the source plane). We find no evidence for a decreasing rotation curve, from which we derive a dynamical mass of $(6.3 \pm 0.7) \times 10^{10} \rm M_{\odot}$ within $r=2.6\pm0.1$~kpc. The discovery of a third, unpredicted, image provides key information for a future improvement of the lensing modeling of the Bullet Cluster and allows a measure of the stellar mass, $1.6^{+1.9}_{-0.86} \times 10^{10} \, M_{\odot}$, unaffected by strong differential magnification. The baryonic mass is is expected to be dominated by the molecular gas content ($f_{gas}\leq80\pm20$\%) based on an $M_{H_2}$ mass estimated from the difference between dynamical and stellar masses. The star formation rate is estimated via the spectral energy distribution ($\rm SFR=190\pm10 M_{\odot} / yr$), implying a molecular gas depletion time of $0.25\pm0.08$\,Gyr.   
\end{abstract}


\section{Introduction} \label{sec:intro}

The flat rotation curves in local spiral galaxies \citep{2001ARA&A..39..137S} are a basic piece of evidence for the existence of dark matter in the Universe. 
At high redshift, studies of rotation curves are limited by the low surface brightnesses of the outskirts of galaxies. 
The recent report of decreasing (as a function of galactocentric radius) rotation curves in massive galaxies at redshift $z=0.9-2.4$ has suggested that dark matter does not dominate the dynamical mass at distances larger than 1.3--1.5 times the galaxy effective radius \citep{2017Natur.543..397G}. 
These findings are influenced by the effect of pressure support in turbulent disks, which is more commonly seen in high-$z$ gas-rich galaxies. 
Probing rotation curves is difficult at high $z$, however this can be aided by strong gravitational lensing \citep{2008Natur.455..775S}.

\begin{figure*}[thb]
\begin{center}
\plotone{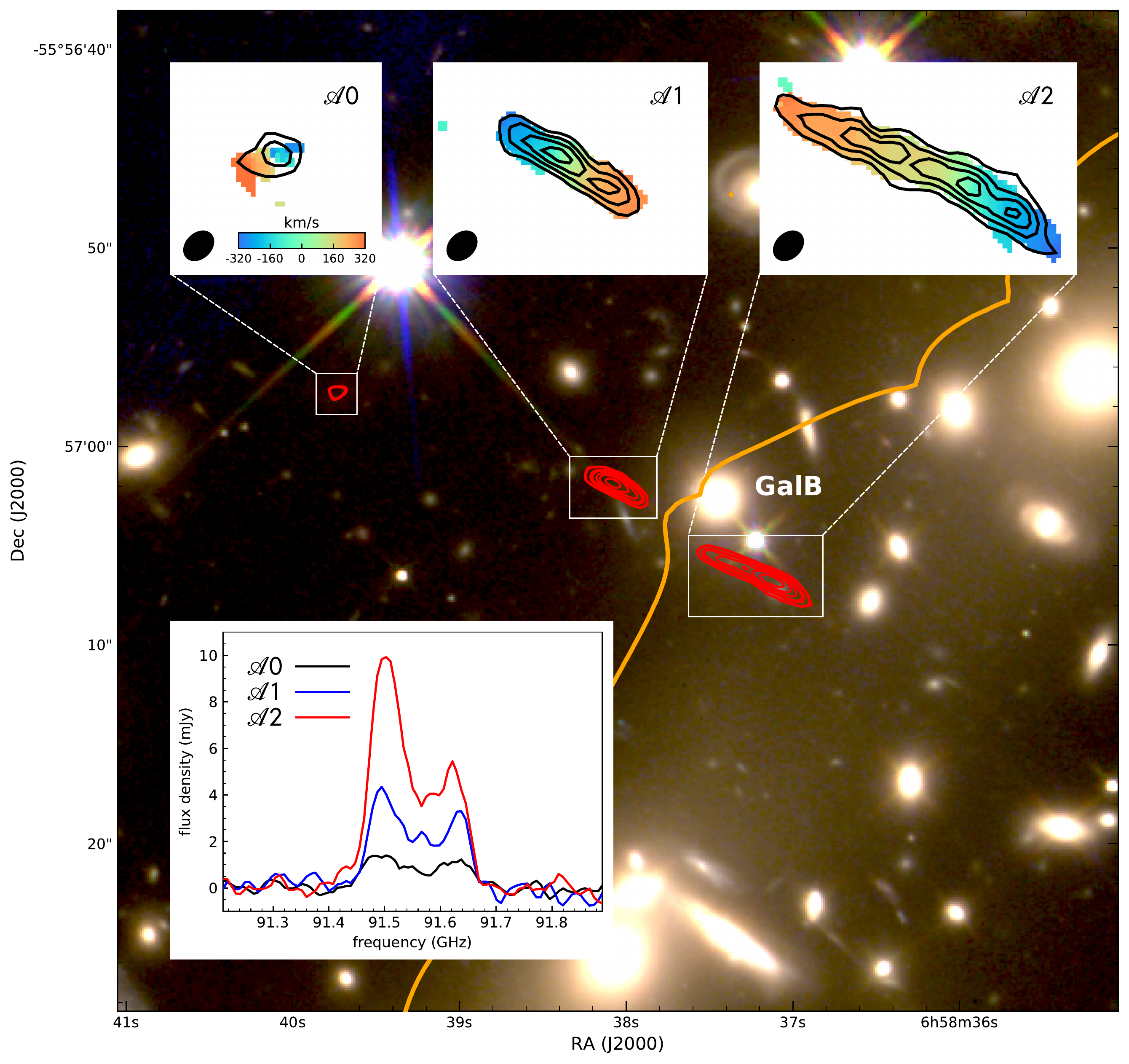}
\caption{HST composite (WFPC2/F814W, WFPC3/F110W and F160W) image and contours (intervals from 20\% to 80\% of 1.16~mJy~beam$^{-1}$ peak intensity) from ALMA band-6 continuum centered at 231~GHz showing the structure of the `Cosmic Seagull' near a bright galaxy (GalB). The orange line represents the  critical line in the lens plane. Inserts at the top exhibit the CO(3--2) rotation velocity map (in colours) and line intensity (in contours in intervals from 20\% to 80\% of 0.49~Jy~beam$^{-1}$~km~s$^{-1}$ peak intensity) for $\mathscr{A}0$, $\mathscr{A}1$, and $\mathscr{A}2$. 
The ALMA synthesized beam (0.5''$\times$0.7'', PA$=43^{\circ}$) is shown as black ellipses. Notice the inverted rotation curves for $\mathscr{A}2$ and  $\mathscr{A}1$, caused by the lensing effect. The central velocity, as well as the velocity range, of $\mathscr{A}0$ is the same as for $\mathscr{A}2$ and $\mathscr{A}1$. 
The insert at the bottom shows the observed spectra from all of the 'Cosmic Seagull' images ($\mathscr{A}0$, $\mathscr{A}1$ and $\mathscr{A}2$).  The spectra have been smoothed to a final resolution of 23.4~MHz (78 km~s$^{-1}$)  to improve visualization. The effect of differential magnification (not corrected in the figure) can be significant in the low-frequency (high-velocity) side of the CO(3--2) emission line, as it approaches the critical line. The similarity of the profiles confirms that $\mathscr{A}0$, $\mathscr{A}1$ and $\mathscr{A}2$ are three images of the same object. \label{co32} }
\end{center}
\end{figure*}

Strong gravitational lensing by massive clusters can boost the signal of background galaxies, offering the opportunity to probe in great detail their internal structures \citep{2010Natur.464..733S}. 
The Bullet Cluster (1ES 0657-558) consists of two merging galaxy clusters at z = 0.296 \citep{1998ApJ...496L...5T}, and its high mass \citep{2006ApJ...652..937B,2016A&A...594A.121P} causes the field to display an exceptionally high number of lensed sub-millimeter galaxies \citep{2010A&A...514A..77J}. 
Here we present the physical properties of galaxy SMM~J0658 at z=2.7793 \citep{2012A&A...543A..62J}, hereafter called the `Cosmic Seagull' (figure \ref{co32}).
This galaxy lies close to a lensing critical line, resulting in one of the largest magnifications reported to date ($\mu$ up to $\sim 50$), which facilitates precise measurements of the kinematics at the outskirts of this dusty, star-forming galaxy.
Through the text we assume a $\Lambda$CDM cosmology with $\Omega_m=0.3$, $\Omega_{\Lambda}=0.7$ and $h=0.7$.

\section{ALMA Observations}

We used the Atacama Large Millimeter/submillimeter Array (ALMA) band-3 to observe the redshifted CO(3--2) line at 91.50~GHz, and also band-6 to target the continuum centered at 231~GHz (344~$\mu$m rest-frame) over a 7.5~GHz bandwidth (project 2015.1.01559.S). 
Calibrated $u$-$v$ products were obtained using the scripts provided to the P.I.\, (version Pipeline-Cycle3-R4-B) within the Common Astronomy Software Application (CASA version 4.7.2).  Observations taken on different days were concatenated together using the task {\sc concat}. 
The continuum in each band was calculated excluding channels presenting line emission, and subtracted from the visibilities with the task {\sc uvcontsub}. 
The imaging was performed with {\sc clean} using a Briggs weighting ({\sc robust}\,$=0.5$), achieving a synthesised beam  of $0.5''\times 0.7''$ for band 3 and $0.5''\times 0.8''$ for band 6. 
The sources were cleaned manually down to a 3\,$\sigma$ level,   using masks at source positions. 
The final CO(3-2) cube has a root mean squared (RMS) noise of 0.2~mJy\,beam$^{-1}$  per 7.8~MHz (26\,km\,s$^{-1}$) channel width in band-3. In the continuum, the noise level is  9$\mu$Jy\,beam$^{-1}$ in band-3 and 25$\mu$Jy\,beam$^{-1}$ in  band-6. 
Moment maps were produced using task {\sc immoments}, excluding  pixels below 4$\sigma$.

Previous observations with the Australia Telescope Compact Array (ATCA) \citep{2012A&A...543A..62J} at $\sim$2'' resolution targeting the CO (1--0) and (3--2) emission described two point-like features. 
Figure \ref{co32} shows that, at high resolution, they appear as two clear arc-like structures ($\mathscr{A}1$, $\mathscr{A}2$) with observed CO(3--2) full extents, at a  level\footnote{20\% of CO(3--2) peak intensity of  0.1~Jy~beam$^{-1}$~km~s$^{-1}$} (lens plane), of $3.4''$ and $6.2''$ respectively  -- comparable to what has been measured in the optical  \citep{2010ApJ...720..245G}. 
The ALMA sub-arcsecond resolution, coupled with the high gravitational magnification, allows us to investigate a region with a full extension of $\sim$\,4.6\,kpc in the source plane with an effective resolution of $\sim 620$~pc
(each pixel corresponds to 0.013~pc in the source plane).  
The data clearly reveal a rotating disk-like structure, with a striking mirror symmetry. 
A third source of emission  is detected ($>5\sigma$) in CO(3--2) with an observed size of $1.2''$ (a deconvolved size of $0.6''$) in the lens plane, sharing the same kinematic characteristics as $\mathscr{A}1$ and $\mathscr{A}2$ but offset to the North-East by 20''. 
This  image was not predicted by any previous lensing model \citep{2009ApJ...706.1201B,2016A&A...594A.121P}. 
We call this component $\mathscr{A}0$.

\section{Methods, Results and Discussion} 

The reconstruction of the source was performed using the LENSTOOL code \citep{2007NJPh....9..447J}, which uses a Bayesian optimization aided by the redshift and location of bright cluster galaxies and arcs, yielding an output for the best estimated parameters. 
We assume that the mass distribution of the Bullet Cluster consists of three components: the intracluster gas mass, the cluster galaxy members, and three dark matter clumps \citep{2009ApJ...706.1201B,2016A&A...594A.121P}. 
To reproduce the morphology of $\mathscr{A}1$ and $\mathscr{A}2$, we employ the best model published to date \citep{2016A&A...594A.121P} (which did not consider $\mathscr{A}0$) but optimizing the parameters of the density profile of a nearby bright galaxy \citep[at $6^h58^m37.449^s$, $-55^{\circ}57'02.59''$, $z=0.2957$,][]{2002A&A...386..816B} seen in the lensing galaxy cluster, by using as constraints the centroid positions of the CO(3--2) rest-frame velocity map.
We fixed the galaxy position (X,Y), adopt a Pseudo-Isothermal Elliptical Mass Distribution  \citep[PIEMD,][]{2005MNRAS.356..309L}, and optimize the remaining parameters with broad uniform priors. 
Given our best-fit parameters (galaxy ellipticity $\epsilon$ = 0.69$\pm$0.09, position angle $\theta$ = 8$^{+18}_{-5}$, characteristic radii $r_{core}$ = 0.06$^{+0.55}_{-0.01}$ kpc, $r_{cut}$ = 63$^{+110}_{-23}$ kpc, velocity dispersion $\sigma_0$ = 88$^{+5}_{-8}$ km\,s$^{-1}$), 
we reproduce the images $\mathscr{A}1$ and $\mathscr{A}2$ with a root-mean-square  error in position in the image plane $rmsi$ = 0.09", and a $\chi^2$ per degree of freedom of $1.2$.

The CO(3--2) surface brightness map (figure \ref{co32}) shows that $\mathscr{A}1$ and $\mathscr{A}2$ are stretched along a  direction almost perpendicular to the critical line,  introducing different magnification factors (differential magnification) over the length of each of the arcs. We estimate that the range of magnification factors are $\mu_{\mathscr{A}1} \approx 11-22$ and $\mu_{\mathscr{A}2} \approx 13-50$ (with an estimated $\pm$10\% error). 

The previously predicted $\mathscr{A}3$ component \citep[identified as K3 by][]{2009ApJ...706.1201B} located $\sim38$'' to the southwest of $\mathscr{A}2$ is at the extreme edge of the ALMA field of view of our observations. 
Making use of archival observations in continuum band-6 (project 2012.1.00261.S, PI J. Richard), we cannot confirm the presence of $\mathscr{A}3$ (at $>5 \sigma$),  given that the local RMS\,$\sim$\,0.3~mJy~beam$^{-1}$ at 230~GHz 
is too high at that position. 
Nevertheless, given the estimated magnification $\mu_{\mathscr{A}3}\sim 5$ predicted by the lens model, we cannot rule out its existence.

\begin{figure}[htb]
\begin{center}
\includegraphics[width=6.5cm]{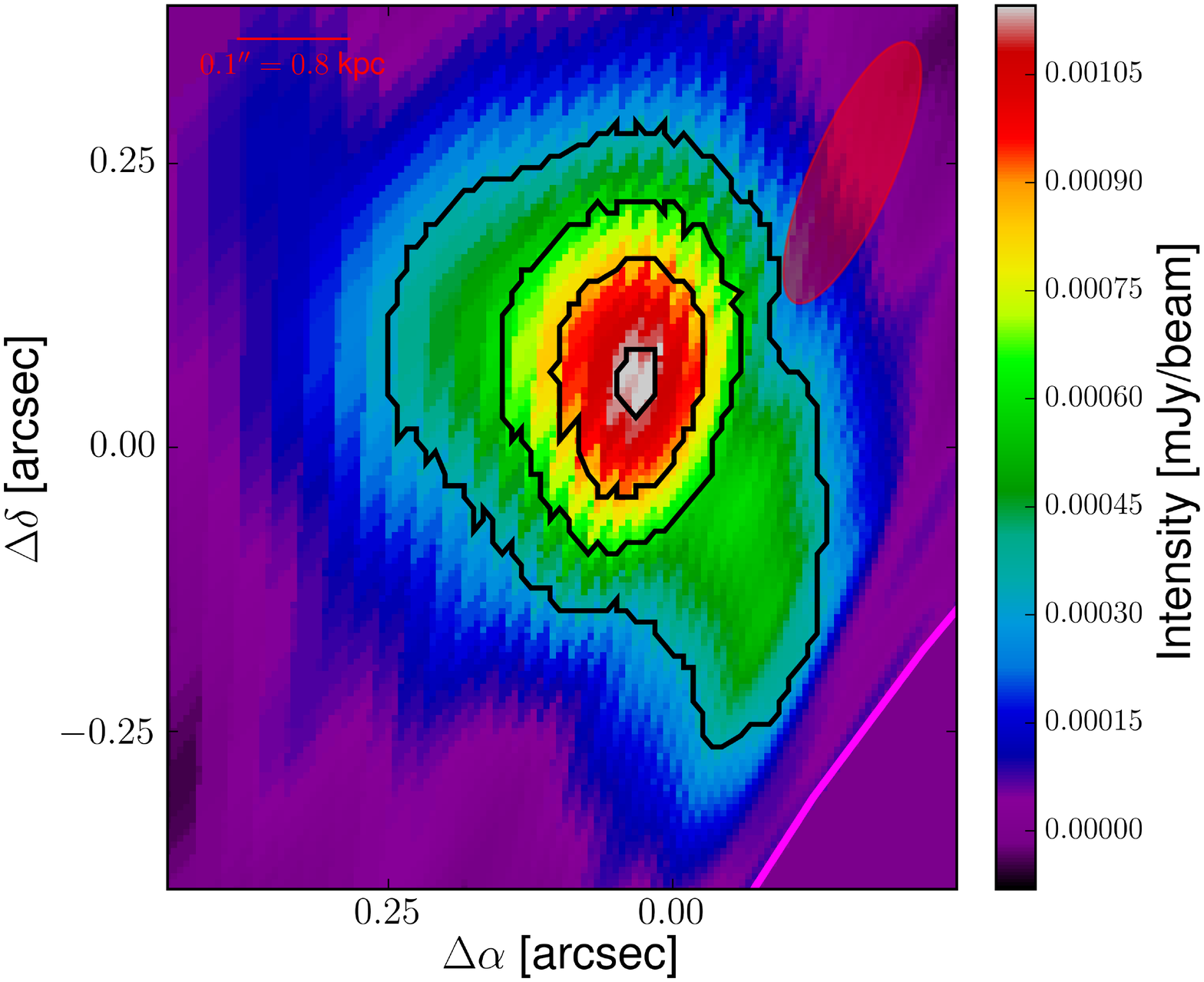}
\includegraphics[width=6.5cm]{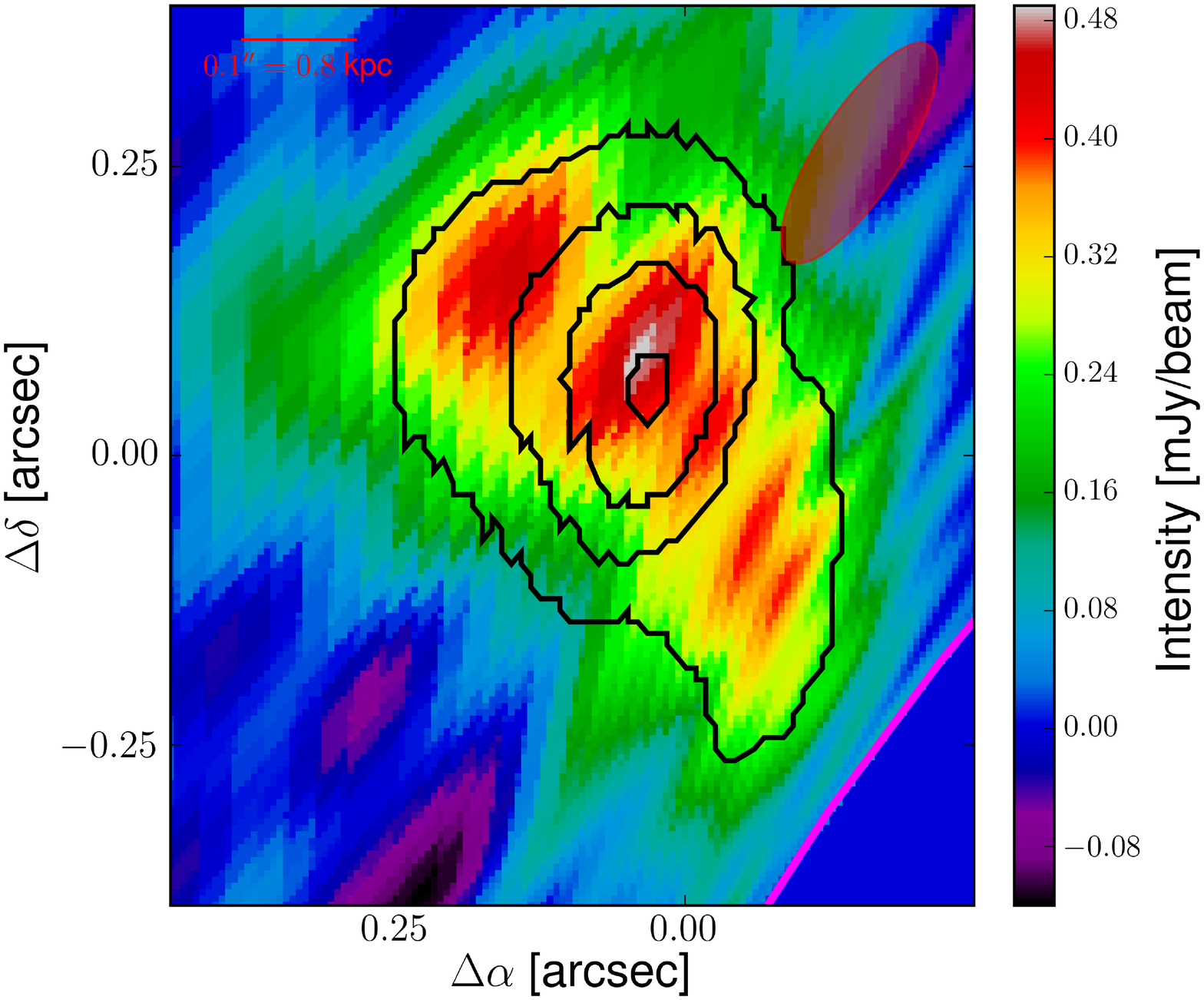}\\
\includegraphics[width=6.5cm]{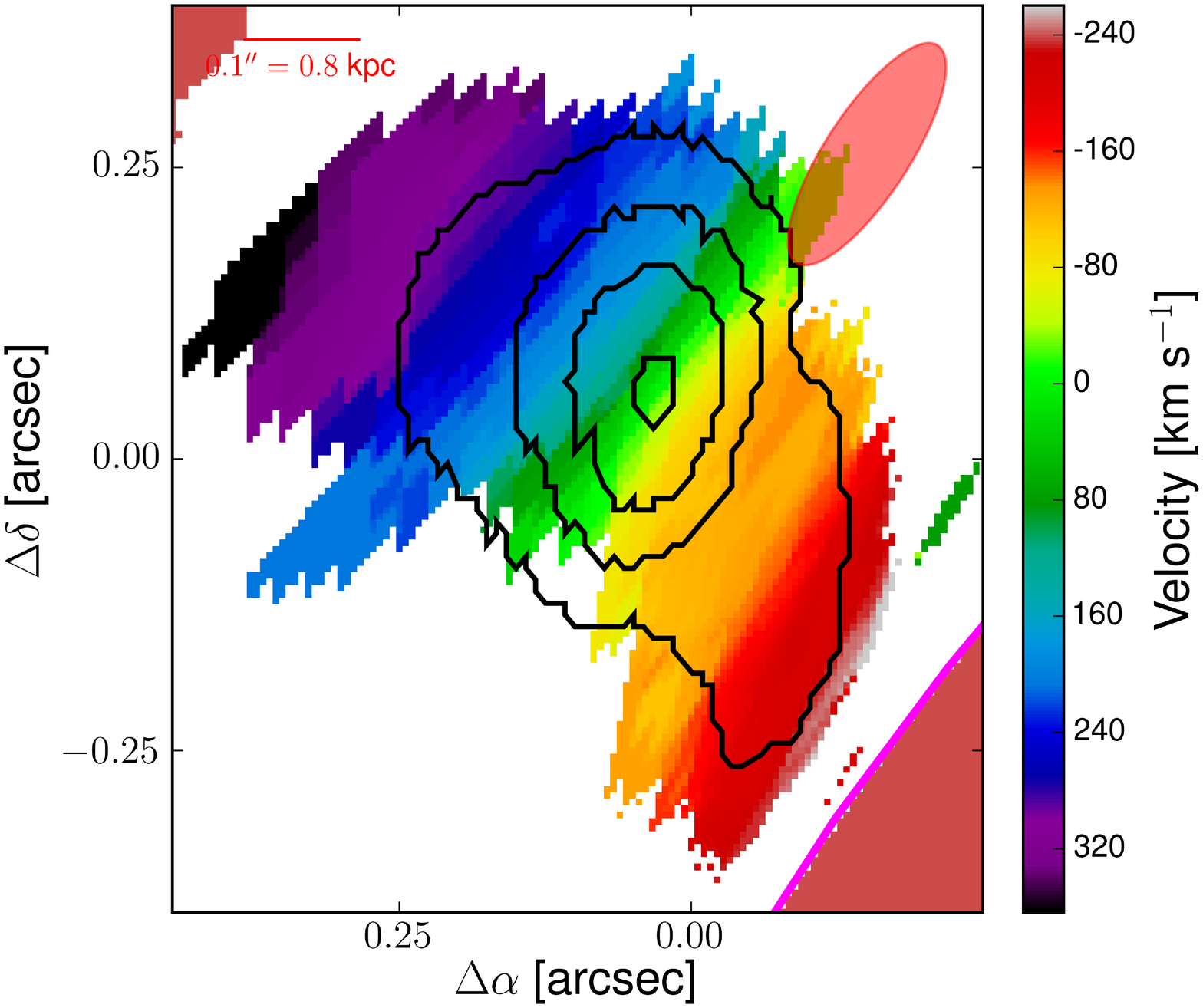}
\includegraphics[width=6.5cm]{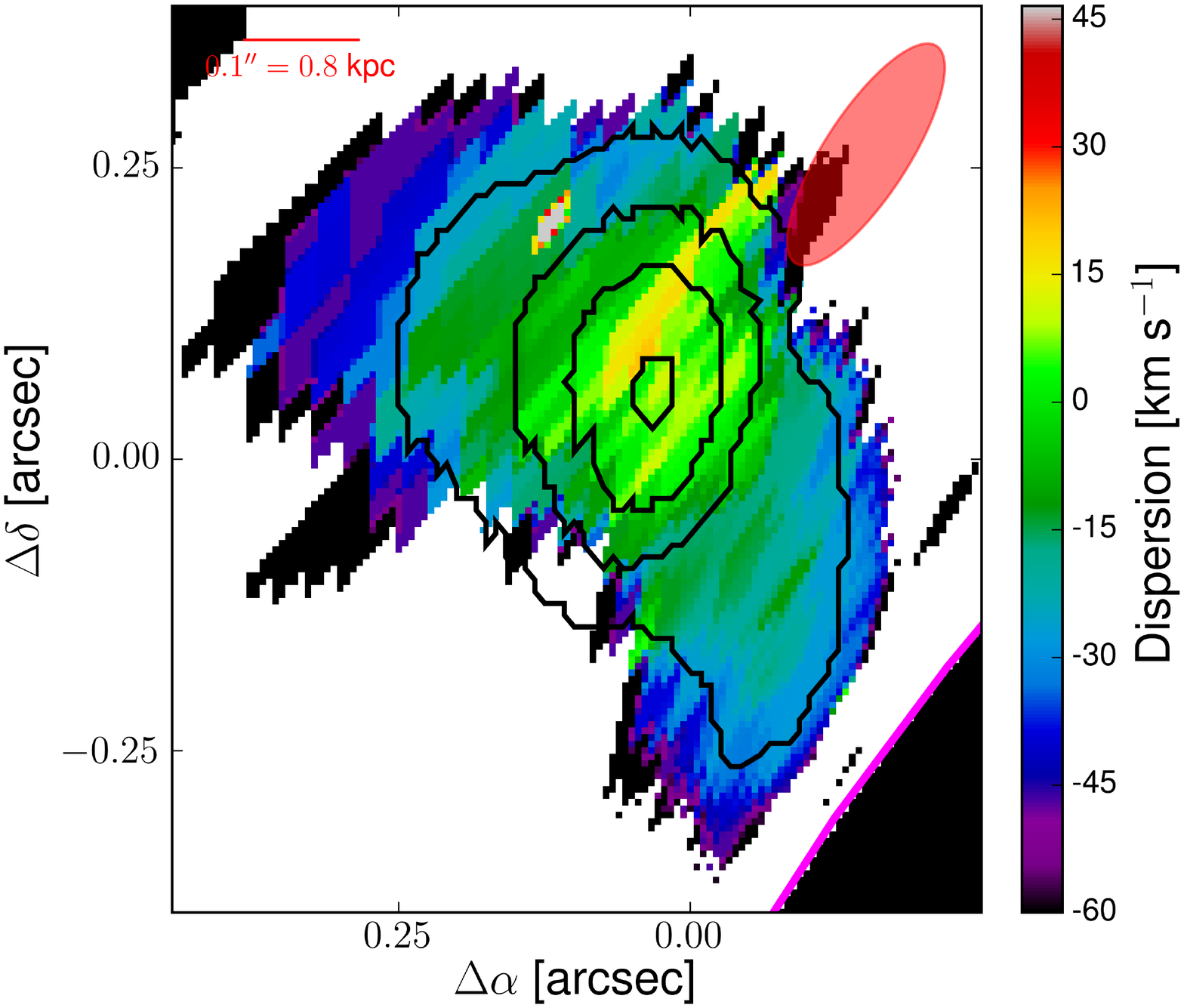}
\end{center}
\caption{Lens model reconstruction of the `Cosmic Seagull' source. The maps show the $\mathscr{A}2$ reconstructed source morphology  for (\textbf{a}) continuum centered at 231~GHz, (\textbf{b}) CO(3--2) surface brightness, 
(\textbf{c}) CO(3-2) rest-frame velocity,
and (\textbf{d}) CO(3--2) velocity dispersion.
The magenta line represents the caustic curve (regions of highest magnification) in the source plane. The black contours show the continuum (intervals from 20\% to 80\% of 1.16~mJy~beam$^{-1}$ peak intensity). The light-red ellipse shows the ALMA synthesized beam in the source plane, providing a mean resolution of $\sim 620$~pc ($\sim0.078''$) along the plane. 
\label{source_rec} }
\end{figure}

We perform a source-plane reconstruction \citep{2012ApJ...746..161S} by ray-tracing the images in the observed frame through the lensing potential of the best-fit lens model, i.e., we `unlensed' the observed images, $\mathscr{A}1$ 
and $\mathscr{A}2$ (Figure \ref{source_rec}), to the source plane. As the best model does not predict $\mathscr{A}0$, this image cannot be 'unlensed'. 
The uncertainty in the magnification map is estimated by constructing a model in which the input values for the seven parameters are chosen as those from the best-fit model $\pm 1 \sigma$ from the $\Delta \chi ^2$ value.  
The resulting magnification maps were compared with the one obtained from the best-fit model, resulting in a 10\% uncertainty.

Figure~\ref{source_rec} shows the reconstructed source corresponding to a rotating disk with a physical radius of 3.3~kpc. 
The increase of the magnification factor towards the caustic curve (i.e., along the arc) produces a decrease in the  effective size of the reconstructed ALMA synthesized beam. 
Considering an average reconstructed synthesized beam
 size along the arc ($0.072''\sim 570$~pc) and the astrometric rms error on the source plane ($0.03''\sim240$~pc), the effective positional error is roughly 620~pc. 
 An interesting discovery is the positional coincidence between the peak of the continuum and the kinematical center of the CO(3--2) velocity dispersion map. 
 There is a smooth velocity gradient (with no secondary peaks) suggesting a stable rotating disk without evidence of significant disruption.

\begin{figure}[htb]
\begin{center}
\includegraphics[width=6cm]{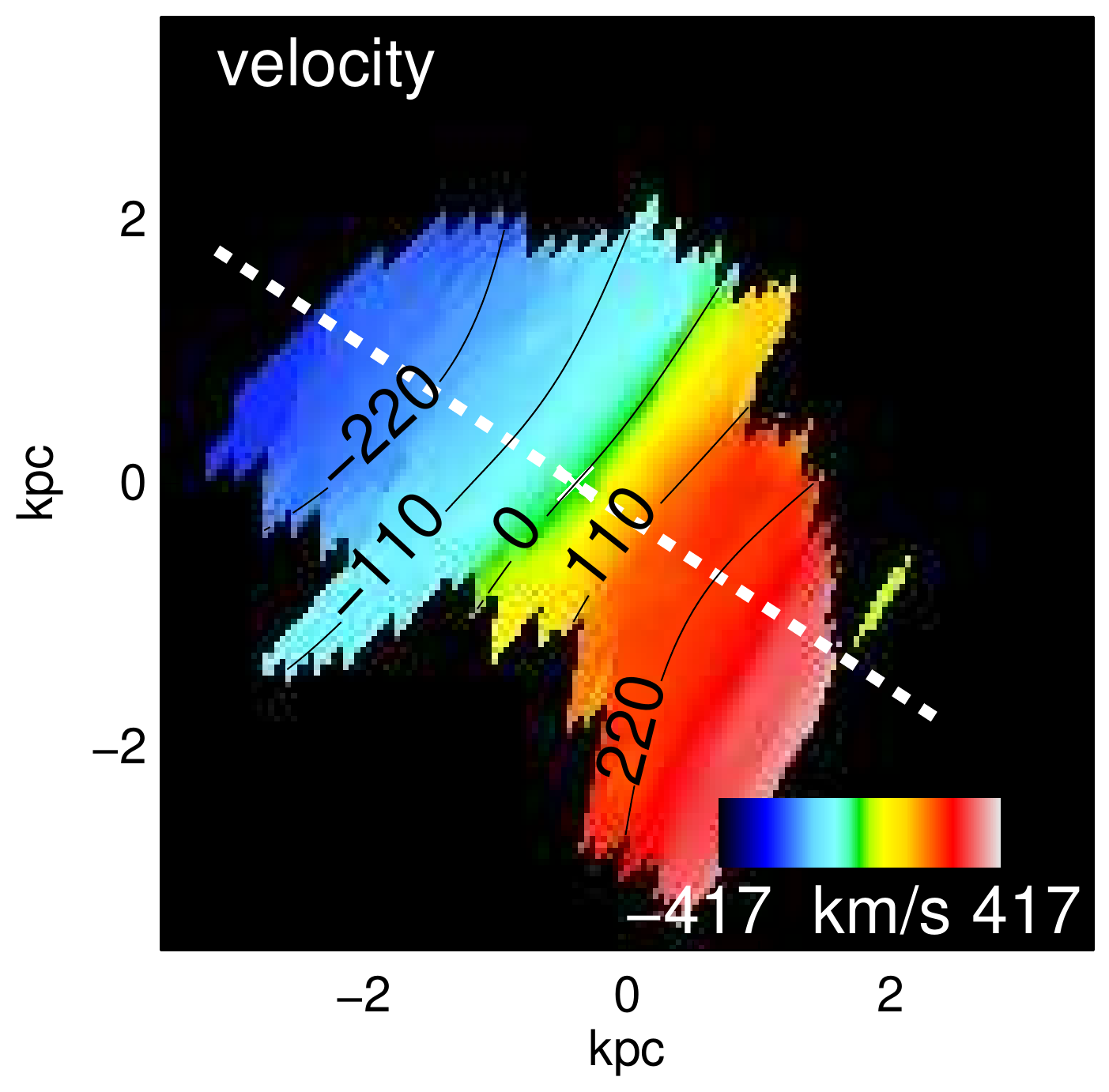}
\includegraphics[width=6.2cm]{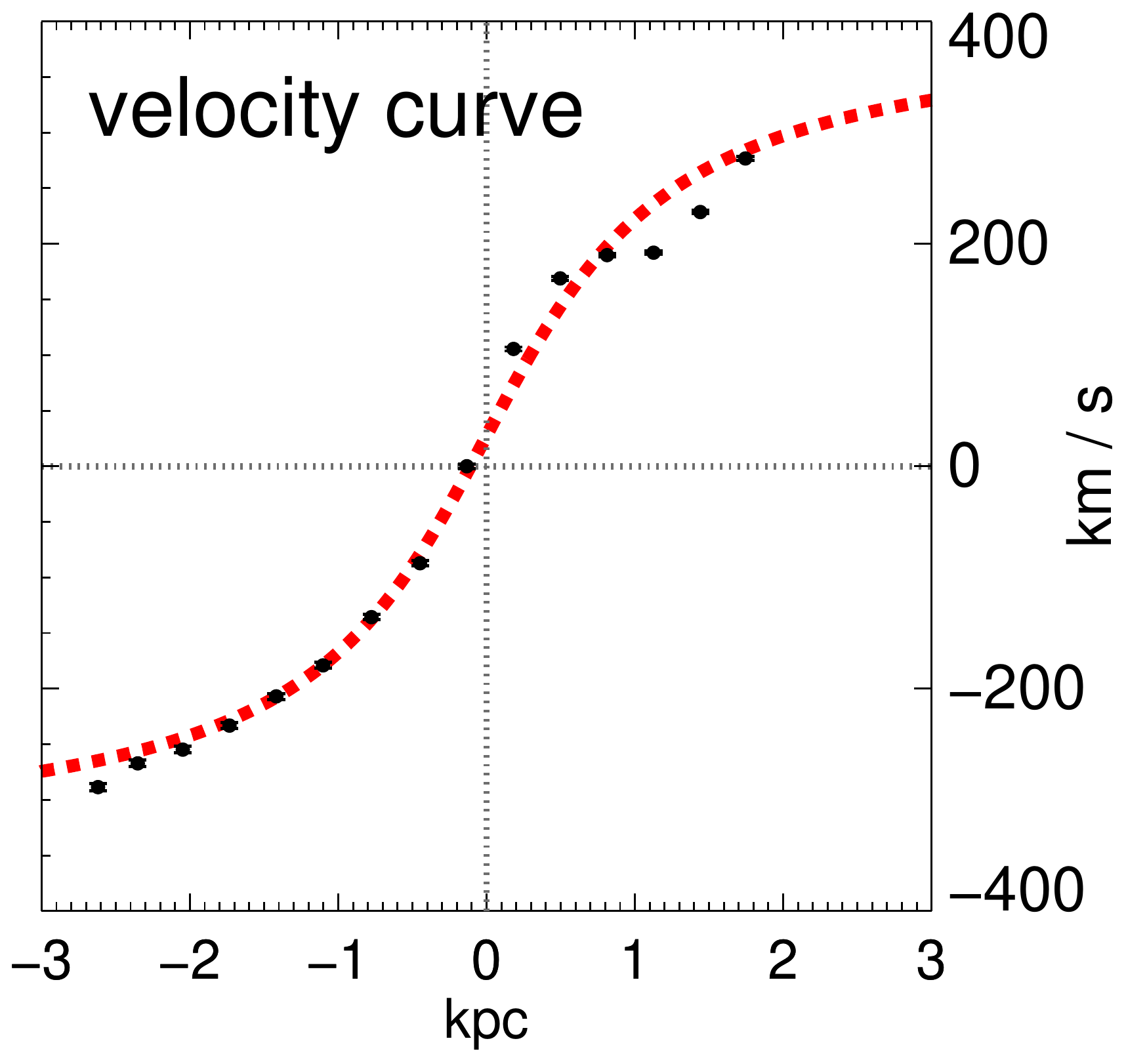}\\
\includegraphics[width=6cm]{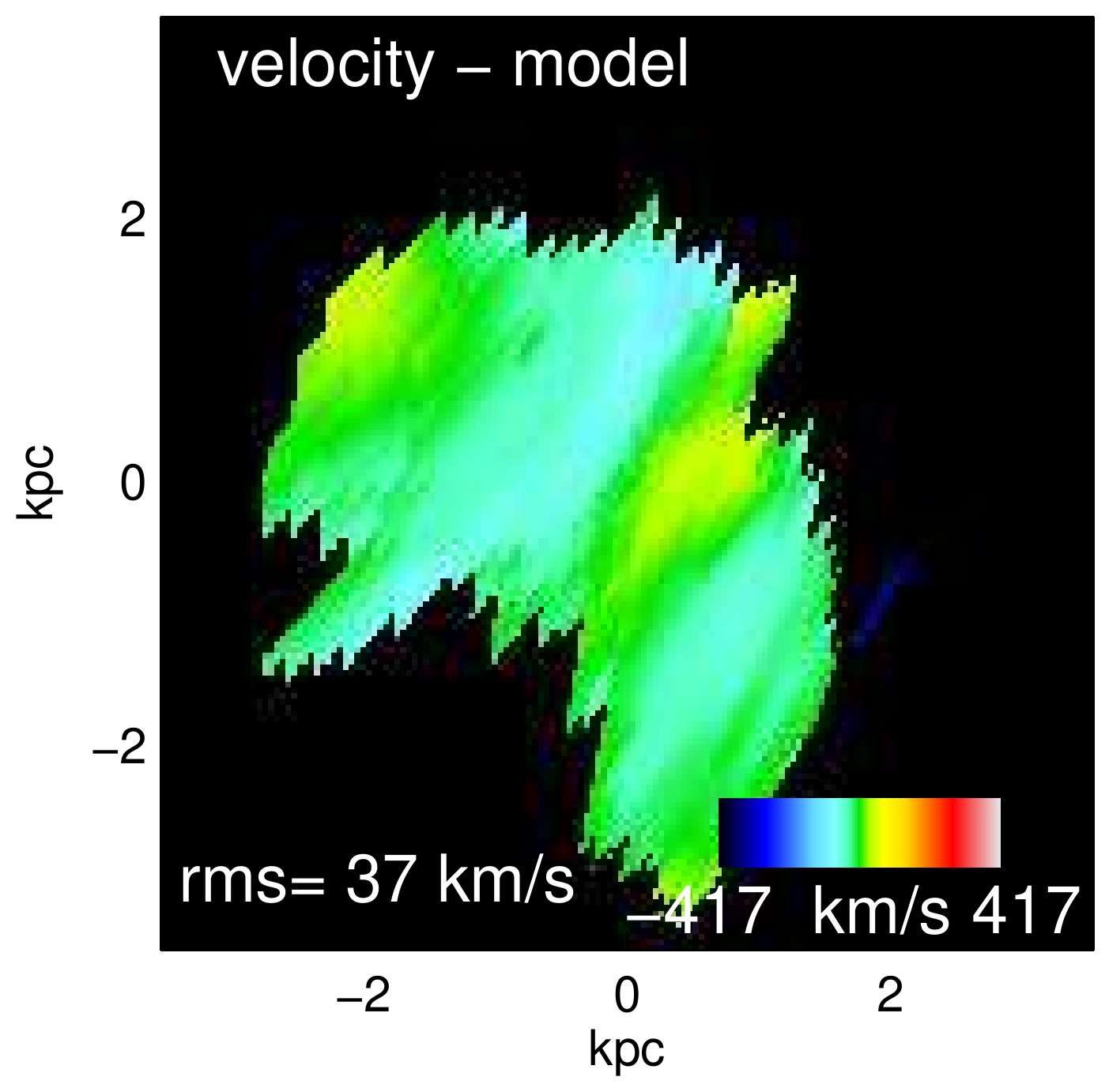}
\includegraphics[width=6.5cm]{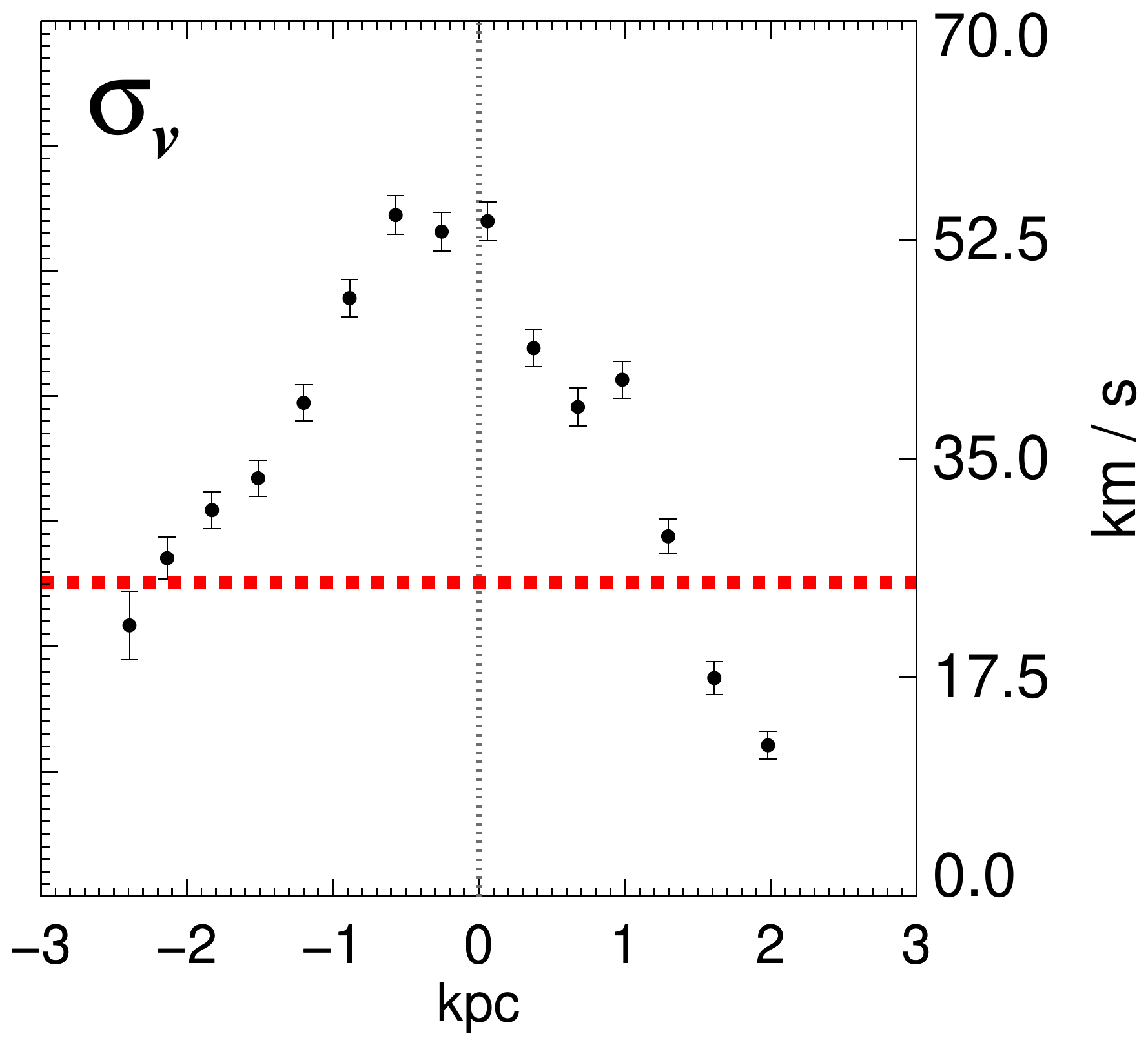}
\end{center}
\caption{Kinematic analysis for the `Cosmic Seagull' reconstructed $\mathscr{A}2$ source using a thick disk model. 
(\textbf{a}) Reconstructed CO(3--2) velocity field with contours from the best fit disk model. The dashed line represents the position angle of the major axis of the galaxy derived from the best two-dimensional kinematic model. (\textbf{b}) Major axis rotation curve for CO(3--2) with $1\, \sigma$ errors (black) and the best fit disk model (red).  (\textbf{c}) Residual map from the best fit disk model. (\textbf{d}) CO(3--2) line width ($\sigma_v$) corrected by the local velocity gradient across the line of sight. The dashed red line represents the mean velocity dispersion.  
\label{kinematicB} }
\end{figure}

To model the kinematic properties of the `Cosmic Seagull', we  adopted as the dynamical center  the location of the peak of the continuum map at rest-frame wavelength 344~$\mu$m. 
Then, we fit the two-dimensional velocity map by assuming 
an arctan velocity profile model\footnote{A Freeman model (exponential disk) was also tried, but this did not provide an improved fit.} ($\rm V(r) = (2/\pi) \, V_{asym} \, arctan (r/r_t)$). 
We adopt a fitting procedure \citep{2012MNRAS.426..935S} to search for four free parameters: the asymptotic rotational velocity $V_{\rm asym}$, the turnover radius $\rm r_t$ \citep{1997AJ....114.2402C}, the position angle $PA$, and the disk inclination $i$. 
We use the best two-dimensional model to construct tilted rings to estimate the radii \citep{2017MNRAS.466..892M} at which the encircled CO(3-2) intensity falls to half its total integrated value ($r_{1/2,\rm CO}$).

According to previous studies of high-$z$ galaxies, we consider a thick disk approximation. 
Although the fitting process was performed for both ($\mathscr{A}1$ and $\mathscr{A}2$) reconstructed sources, the high magnification in the external part of the rotation curve in $\mathscr{A}2$ increases the signal-to-noise ratio, allowing us to extend the analysis further in radius than for $\mathscr{A}1$. Thus, in the following we quote the results obtained for $\mathscr{A}2$ analysis. We found a best inclination of $i=(59\pm 2)^{\circ}$  and a rotational velocity corrected for inclination $V_{\rm rot}=(323 \pm 8)$~km~s$^{-1}$ \citep[defined as the velocity observed at 2.6~kpc, e.g.][]{2017MNRAS.471.1280T}.  
This implies a total dynamical mass of $M_{dyn} \sin ^2 i =(4.6\pm0.7) \times 10^{10} \rm M_{\odot}$ inside $r=2.6 \pm 0.1$~kpc, which is in rough agreement with the former results ($\sim 3.4\times 10^{10} \rm M_{\odot}$ using a size $L=2.6 \rm kpc$) derived simply from the CO emission line widths \citep{2012A&A...543A..62J}. 
Our sensitive ALMA observations show a double-peaked integrated CO(3--2) line profile (with apparent widths of 220-250~km~s$^{-1}$ FWHM), but differentially magnified by approximately a factor of 2  across $\mathscr{A}2$, especially at the low-frequency edge of the line (see figure \ref{co32}). 
The lower signal-to-noise of the previous work \citep{2012A&A...543A..62J}  limited the accuracy of the line widths on which their simple estimate of dynamical mass was based. 
The dynamical mass corrected by inclination is $(6.3\pm0.7) \times 10^{10} \rm M_{\odot}$ inside $2.6 \pm 0.1$~kpc.

The rotation curve (Figure~\ref{kinematicB}) also shows the rotational velocity increases 
as a function of galactocentric radius as far as our observations extend (2.5~kpc $\sim 1.6 r_{1/2,CO}$).  
Recent studies using the H$\alpha$ emission line \citep{2017Natur.543..397G} have suggested that massive high-redshift galaxies ($z\sim 0.9-2.4$) exhibit a decrease in rotational velocity for radii larger than $\approx (1.3-1.5) r_{1/2}$, suggesting only minor contribution by dark matter at the galaxy outskirts. Instead, Figure~\ref{kinematicB} suggests that the `Cosmic Seagull', a disk-like galaxy  at high-$z$,  shows no evidence of such decrease in the rotation velocity \citep[as also found in other galaxies, e.g.][]{2018arXiv180704291X}.

\begin{figure}[htb]
\plotone{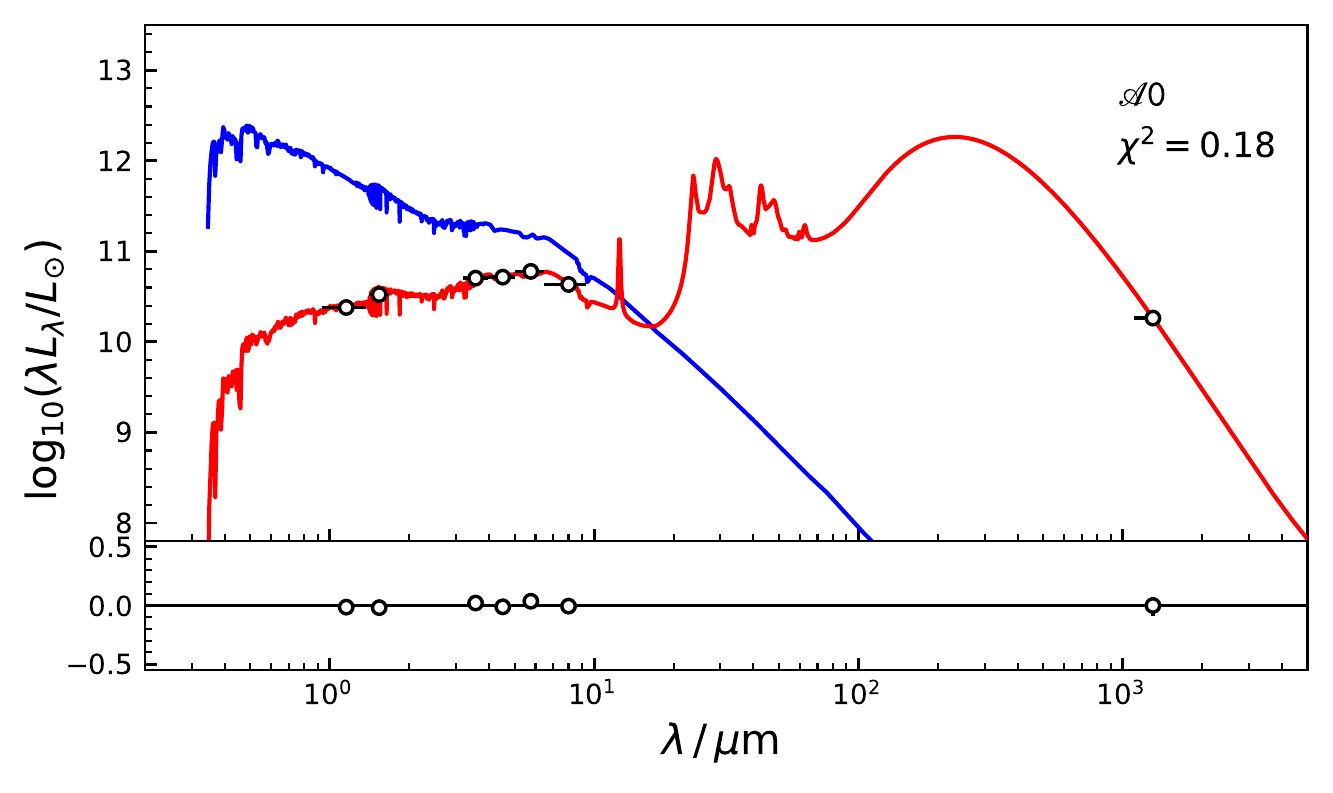} 
\caption{Observed spectral energy distribution for $\mathscr{A}0$ using HST/WFC3 F110W, F160W bands, IRAC 3.6, 4.5, 5.8 and 8.0~$\mu$m bands, and ALMA continuum centered at 231~GHz (1.3mm) and $\mu_{\mathscr{A}0}=4$ at $z=2.7779$. We plot the best MAGPHYS SED fit comprising the unattenuated stellar continuum (blue line) and the observed (reprocessed light) emission including the cold dust emission (red line). Errors are included in quadrature with a 10\,\% error in the magnification. Continuum in band-3 is not included as it is not detected at $\mathscr{A}0$.  \label{sedA0} }
\end{figure}

A previous estimate of the stellar mass of SMM~J0658 made use of $\mathscr{A}1$ and $\mathscr{A}2$ infrared data \citep{2010ApJ...720..245G}. 
Unfortunately, $\mathscr{A}1$ and $\mathscr{A}2$ are projected near one of the brightest elliptical galaxies in the field (figure \ref{co32}), thus this galaxy flux needs to be modeled and subtracted to obtain  uncontaminated fluxes from the arcs. 
The magnification gradient along the arcs also complicates the unlensed flux calculation. 
In this work, we take advantage of the third lensed image, $\mathscr{A}0$, to estimate the stellar mass. 
In particular, this image is preferred  as it is not significantly affected by near neighbors, so  can provide  a clean estimate for the stellar mass via spectral energy distribution fitting (SED; see Fig.~\ref{sedA0}). 
Based on the lensing model, this image is far from the lens critical line and it is expected to have a magnification of $\mu = 4$ ($\pm$10\% error). 
This magnification is also consistent with the CO(3--2) line flux when compared with $\mathscr{A}1$ and $\mathscr{A}2$ fluxes. 
We calculate the $\mathscr{A}0$ SED by using aperture photometry in data from HST/WFC3 (F110W and F160W), Spitzer/IRAC 3.6, 4.5, 5.8 and 8.0~$\mu$m bands, and ALMA continuum in band-6. 
Using the MAGPHYS Bayesian SED fitting code  \citep{2008MNRAS.388.1595D,2015ApJ...806..110D} we find a total stellar mass of $M_{star}=1.6^{+1.9}_{-0.86} \times 10^{10} \, M_{\odot}$ within a 2.1~kpc lens-corrected aperture  (i.e., $\rm 4.2~kpc/|1-\kappa|\simeq 4.2~kpc/\sqrt{\mu}$ considering that the shear is $\simeq0$ at $\mathscr{A}0$ position), consistent with previous results within the errors \citep[$1.3 \times 10^{10} M_{\odot}$ using an average magnification of 30 for $\mathscr{A}1$ and $\mathscr{A}2$,][]{2010ApJ...720..245G}.    
The star formation rate (SFR) from SED fitting is estimated at (190$\pm$10) \,M$_\odot$\,yr$^{-1}$, implying a specific SFR of 
sSFR\,$=12^{+14}_{-7}$~Gyr$^{-1}$, in rough agreement with that estimated for normal 'main sequence' galaxies at $z\sim2.8$ \citep{2015ApJ...800...20G}. 
This result highlights the importance of gravitational magnification for extracting details of those normal galaxies which are usually beneath the noise levels but compose the bulk of the population at high-$z$.

Making use of the CO(3--2) emission line to estimate the molecular gas content \citep{2013ARA&A..51..207B} in the `Cosmic Seagull' and following a similar approach as before, in $\mathscr{A}0$  we find a CO(3--2) velocity integrated flux density of $228\pm44$\,mJy\,km\,s$^{-1}$ (corrected by a magnification $\mu=4$, at 91.5~GHz).  
Using a classical \citep{2005ARA&A..43..677S} conversion to get the CO(3--2) luminosity, $L'_{\rm CO(3-2)}$, from observed velocity integrated flux densities, a typical ratio \citep{2013ARA&A..51..105C} of $L'_{\rm CO(3-2)}/L'_{\rm CO(1-0)}=0.56$, and an upper limit for $\alpha_{\rm CO}$ of $3.0\,{\rm M_\odot\,(K\,km\,s^{-1}\,pc^2)^{-1}}$ derived from the dynamical mass ($M_{H_2}\leq M_{dyn}-M_{star}$), we estimate a  maximum molecular gas content of $M_{\rm H_2} = (4.7\pm1.4)\times10^{10}\,M_\odot$.  
Considering the estimated dust mass from the MAGPHYS fit, 
$M_{\rm dust}=(1.86\pm 0.07)\times10^8 {\rm M}_{\odot}$, this results in a maximum molecular gas-to-dust mass ratio of $M_{H_2}/M_{dust}=300\pm80$. 
The corresponding maximum molecular gas fraction  $f_{\rm gas}=M_{\rm H_2}/(M_{\rm H_2} + M_{star})\simeq80\pm20\%$  indicates that baryonic mass is most probably dominated by the molecular gas as commonly found in normal high-$z$ galaxies \citep{2010Natur.463..781T,2010ApJ...713..686D} unless $\rm \alpha_{CO}$ is reduced to $0.8\,{\rm M_\odot\,(K\,km\,s^{-1}\,pc^2)^{-1}}$, a value typical of disturbed ULIRG-like galaxies. 
Compared to the previous estimates obtained with the ATCA telescope \citep{2012A&A...543A..62J} which combined $\mathscr{A}1$ and $\mathscr{A}2$ fluxes, after scaling to the same $\alpha_{\rm CO}$ and assuming an averaged $\mu=$\,30 at those positions, we find that our results for $f_{gas}$ are a factor of two higher, 
emphasizing the  power of our new ALMA observations given the uncertainties introduced by differential magnification in previous $\mathscr{A}1$ and $\mathscr{A}2$ images.

Assuming a constant SFR, we estimate a molecular gas depletion time of $\geq 0.25\pm0.08$\,Gyr, in agreement with those commonly seen in rotating disk galaxies.

\section{Conclusions}
These new ALMA observations reveal for the first time an exquisite system with multiple images behind the Bullet Cluster (the `Cosmic Seagull'). 
The kinematic information at the outskirts of a galaxy at $z\simeq2.8$ is facilitated by one of the largest gravitational magnifications ever recorded ($\mu \lesssim 50$).  
The strong lensing of this system reveals its internal dynamics, stellar mass, and gas distribution, providing an excellent opportunity to explore the internal structure of a normal (following the 'main sequence') high-z galaxy which would otherwise be undetectably faint.
This exceptionally magnified object observed at sub-arcsecond resolution provides us with a detailed rotational curve near the peak of the galaxy mass assembly epoch. 
Our ALMA data have shown that SMM~J0658 is a disk-like galaxy, with a dynamical mass of $(6.3\pm0.7) \times 10^{10} \rm M_{\odot}$ inside $r=2.6 \pm 0.1$~kpc, where most of its baryonic mass is probably dominated by the molecular gas content ($f_{gas}\leq80\pm20$\%). 
The correspondence between the peak of the cold dust emission and the center of the velocity dispersion profile shows that this galaxy is not involved in a major merging event, supported also by the relatively long molecular gas depletion time, $\geq 0.25\pm0.08$\,Gyr.

\acknowledgments

The authors gratefully acknowledge the anonymous referee for thoughtful comments, J. Richard and D. Paraficz for the access to their best-fit model parameters, M. Swinbank for facilitating the access to his dynamical modelling code, and R. Genzel for valuable comments.
 V.M., T.V., and J. Maga\~na gratefully acknowledge support from the  PROGRAMA UNAM-DGAPA-PAPIIT IA102517.  E.I.\ acknowledges partial support from FONDECYT through grant N$^\circ$\,1171710. 
 J. Molina acknowledges support from CONICYT Chile (CONICYT-PCHA/Doctorado-Nacional 2014-21140483).
T.M.H. acknowledges the support from the Chinese Academy of Sciences (CAS) and the National Commission for Scientific and Technological Research of Chile (CONICYT) through a CAS-CONICYT Joint Postdoctoral Fellowship administered by the CAS South America Center for Astronomy (CASSACA) in Santiago, Chile. 
M.B. acknowledges support from STFC. 
 D.G. acknowledges the support from CONICYT Astronomy Program, project ALMA-CONICYT Support Astronomer Fund 31AS00002. 
 J. Maga\~na acknowledges the support from CONICYT/FONDECYT project 3160674.
This paper makes use of the following ALMA data: ADS/JAO.ALMA \#2012.1.00261.S,   ADS/JAO.ALMA \#2015.1.01559.S.  ALMA is a partnership of ESO (representing its member states), NSF (USA) and NINS (Japan), together with NRC (Canada), MOST and ASIAA (Taiwan), and KASI (Republic of Korea), in cooperation with the Republic of Chile. The Joint ALMA Observatory is operated by ESO, AUI/NRAO and NAOJ.

\vspace{5mm}

\end{document}